\documentclass[floats,preprintnumbers,groupedaddress,superscriptaddress,floatfix]{revtex4}
\usepackage{amsmath,amssymb,epsfig,color,bbold,ulem}
\allowdisplaybreaks

\newcommand{\nl}{\nonumber \\ }

\newcommand{\x}{x}
\newcommand{\y}{y}

\begin{document}

\preprint{FERMILAB-PUB-21-232-T, LA-UR-21-27844, USTC-ICTS/PCFT-21-32}

\title{QED radiative corrections for accelerator neutrinos}

\author{Oleksandr Tomalak}
\email{Corresponding Author. Email: tomalak@lanl.gov}
\affiliation{University of Kentucky, Department of Physics and Astronomy, Lexington, KY 40506, USA}
\affiliation{Fermilab, Theoretical Physics Department, Batavia, IL 60510, USA}
\affiliation{Theoretical Division, Los Alamos National Laboratory, Los Alamos, NM 87545, USA}

\author{Qing Chen}
\affiliation{University of Kentucky, Department of Physics and Astronomy, Lexington, KY 40506, USA}
\affiliation{Interdisciplinary Center for Theoretical Study, University of Science and Technology of China, Hefei, Anhui 230026, China}
\affiliation{Peng Huanwu Center for Fundamental Theory, Hefei, Anhui 230026, China}

\author{Richard J. Hill}
\affiliation{University of Kentucky, Department of Physics and Astronomy, Lexington, KY 40506, USA}
\affiliation{Fermilab, Theoretical Physics Department, Batavia, IL 60510, USA}

\author{Kevin S. McFarland}
\affiliation{University of Rochester, Department of Physics and Astronomy, Rochester, NY 14627, USA}

\maketitle

\section*{ABSTRACT}
Neutrino oscillation experiments at accelerator energies aim to establish charge-parity violation in the neutrino sector by measuring the energy-dependent rate of $\nu_e$ appearance and $\nu_\mu$ disappearance in a $\nu_\mu$ beam. These experiments can precisely measure $\nu_\mu$ cross sections at near detectors, but $\nu_e$ cross sections are poorly constrained and require theoretical inputs. In particular, quantum electrodynamics radiative corrections are different for electrons and muons. These corrections are proportional to the small quantum electrodynamics coupling $\alpha \approx 1/137$; however, the large separation of scales between the neutrino energy and the proton mass ($\sim{\rm GeV}$), and the electron mass and soft-photon detection thresholds ($\sim{\rm MeV}$) introduces large logarithms in the perturbative expansion. The resulting flavor differences exceed the percent-level experimental precision and depend on nonperturbative hadronic structure. We establish a factorization theorem for exclusive charged-current (anti)neutrino scattering cross sections representing them as a product of two factors. The first factor is flavor universal; it depends on hadronic and nuclear structure and can be constrained by high-statistics $\nu_\mu$ data. The second factor is non-universal and contains logarithmic enhancements, but can be calculated exactly in perturbation theory. For charged-current elastic scattering, we demonstrate the cancellation of uncertainties in the predicted ratio of $\nu_e$ and $\nu_\mu$ cross sections. We point out the potential impact of non-collinear energetic photons and the distortion of the visible lepton spectra, and provide precise predictions for inclusive observables.

\section*{INTRODUCTION}

Current and future accelerator neutrino oscillation experiments~\cite{Abe:2011ks,Abe:2019vii,Ayres:2007tu,Acero:2019ksn,Abi:2020evt,Abe:2015zbg} observe primarily muon neutrinos and antineutrinos in their near detectors, but must precisely interpret electron-neutrino and antineutrino interactions in far detectors to measure oscillation probabilities. Over much of the available parameter space, the discovery of CP violation at next-generation experiments will require as-yet unachieved percent-level control over $\nu_e$ appearance signals~\cite{NuSTEC:2017hzk,DUNE:2020ypp}. Therefore, the precise calculation of differences between muon- and electron-neutrino interactions, including QED radiative corrections, is a critical input to current and future experiments. In this work, we describe a computational framework for these calculations and present results for the basic (anti)neutrino-nucleon charged-current elastic scattering process. We show how important flavor ratios are insensitive to uncertain hadronic and nuclear parameters, so that our results can be applied to experiments with nuclear targets.

\section*{RESULTS}

\subsection*{Factorization}

\begin{figure}[tb]
 \centering
 \includegraphics[height=0.25\textwidth]{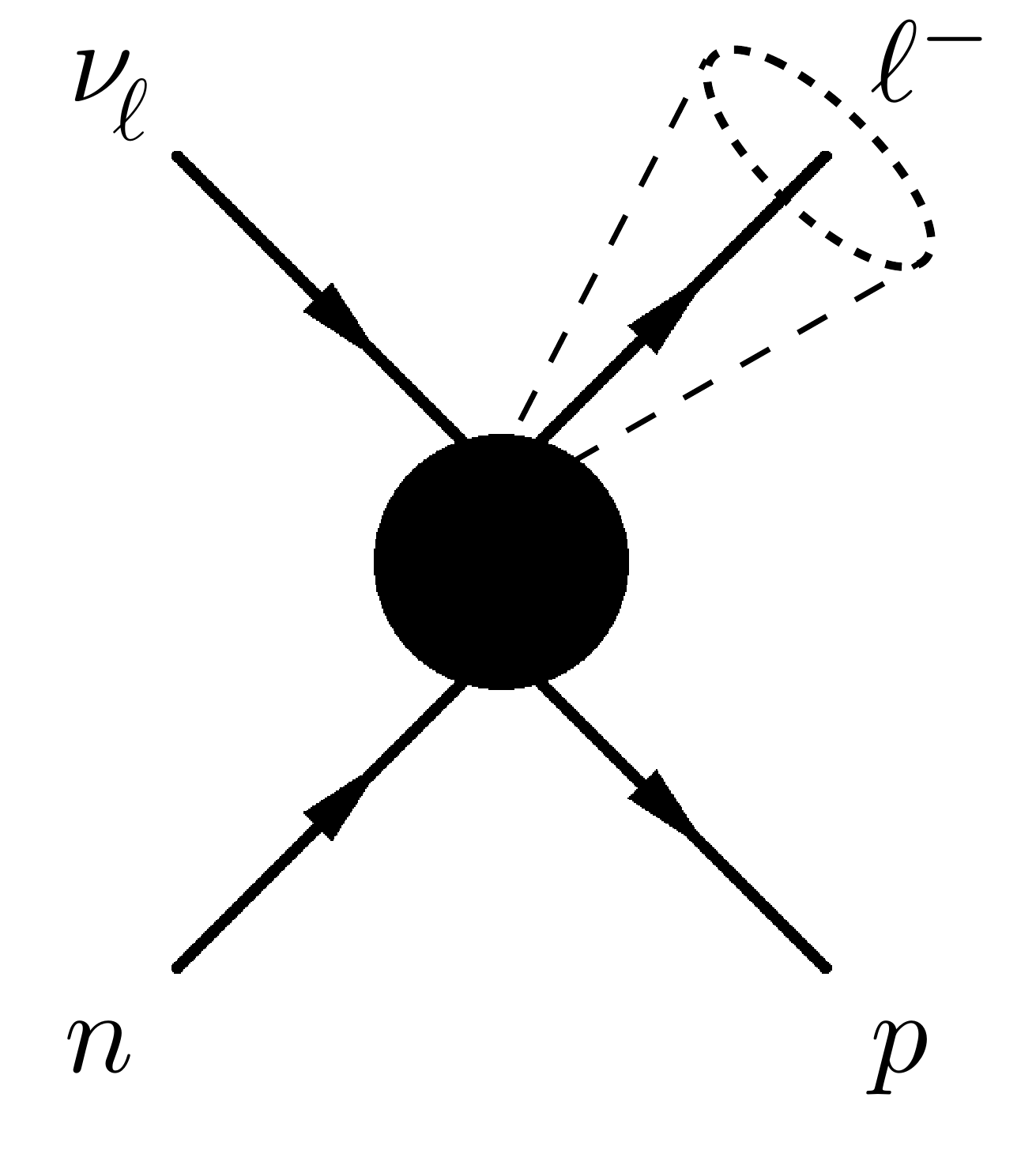}  \caption{Schematic representation of charged-current elastic event. Photons that are within an angle $\Delta\theta$ of the charged lepton, or that have energy below $\Delta E$, are included in the cross section. \label{fig:CCQE}}
 \end{figure}

The separation of scales between the large neutrino energy, the smaller lepton masses, and the soft-photon detection thresholds allows us to apply powerful effective field theory techniques to neutrino scattering. In particular, soft-collinear effective theory (SCET)~\cite{Bauer:2000ew,Bauer:2000yr,Bauer:2001ct,Bauer:2001yt,Chay:2002vy,Beneke:2002ph,Hill:2002vw,Becher:2014oda,Hill:2016gdf} establishes the following factorization theorem for the charged-current elastic process depicted in Fig.~\ref{fig:CCQE}:
\begin{equation}
 \frac{\mathrm{d}\sigma}{ \mathrm{d} Q^2 \mathrm{d} x} = H\left( \frac{\mu}{\Lambda}\right) J\left( \frac{\mu}{m_\ell}, {\Delta \theta \, \frac{E_\ell}{m_\ell}}, x\right) S\left( \frac{\mu}{\Delta E}, x \right) \,. \label{eq:factorization_formula}
\end{equation}
Here $x$ denotes the ratio of  the charged lepton energy $E_\ell$ to the total energy of lepton and photon, $m_\ell$ and $E_\ell$ are the charged lepton mass and energy, and $\mu$ is the renormalization scale. We integrate Eq.~(\ref{eq:factorization_formula}) over the variable $x$ evaluating all observables in this paper. The hard scale is $\Lambda \sim M \sim E_\nu \sim Q$, where $M$ is the nucleon mass and $Q^2$ denotes the momentum transfer between initial and final nucleons. The quantities $\Delta E$ and $\Delta \theta$ denote soft energy and angular acceptance parameters that we specify below. An analogous factorization theorem for elastic electron-proton scattering was presented in Ref.~\cite{Hill:2016gdf}. The charged-current (anti)neutrino-nucleon process differs in that: 1) the electric charges of external particles are different; 2) the underlying quark-level process is weak versus electromagnetic; and 3) real collinear photon radiation is included for typical neutrino detectors.  These differences are reflected in different soft, hard, and jet functions, respectively, compared to the electron-proton scattering case. The soft and jet functions are trivial at the leading order, $S=1$ and $J=\delta(1-x)$, and higher orders can be computed in perturbation theory~\cite{Burgers:1985qg,Korchemsky:1987wg,Kniehl:1989kz,Kilian:1992tj,Hoang:1995ex,Mastrolia:2003yz,Bernreuther:2004ih,Becher:2007cu,Becher:2009qa,Becher:2009kw,Arbuzov:2015vba,Hill:2016gdf,Tomalak:2019ibg}. The hard function contains hadronic physics~\cite{Galster:1971kv,LlewellynSmith:1971uhs,Lepage:1980fj,Kopecky:1995zz,Kopecky:1997rw,Kelly:2004hm} and is nonperturbative. At leading order, it is expressed in terms of nucleon form factors~\cite{LlewellynSmith:1971uhs}. We summarize the explicit components of the factorization theorem through one-loop order in the Methods section. Further details are provided in Ref.~\cite{Tomalak:2022xup}.

In neutrino detectors, photons are spatially localized when they are sufficiently energetic that $e^+e^-$ pair production is their dominant scattering mechanism.   
We identify $\Delta E$ as the minimum energy for this to occur, i.e. photons with energy below $\Delta E$ are not seen by the detector. A photon with energy above $\Delta E$ 
will be absorbed into the reconstructed electron if the photon's direction with respect to the electron is within the angular size $\Delta \theta$  of the electron's shower in the detector. We discuss the determination of $\Delta E$ and $\Delta \theta$ for illustrative cases in the Methods (Photon energy cutoff and angular resolution parameters) section.

The effective theory is constructed as an expansion in powers of the small parameter $\lambda \sim \Delta E/\Lambda$. The lepton mass satisfies $m_\ell \lesssim \sqrt{\lambda}\Lambda$, and the jet angular resolution satisfies $\Delta \theta \lesssim \sqrt{\lambda}$. For the T2K/HyperK, NOvA, and DUNE experiments, appropriate choices are $\Delta E\sim {\rm few}\times 10\,{\rm MeV}$ and $\Delta \theta \lesssim 10^\circ$, and therefore these conditions are satisfied with the power counting parameter $\lambda$ at the percent level. The factorization formula is valid up to power corrections of relative size ${\cal O}(\lambda)$. For numerical evaluations, we include the complete lepton-mass dependence for tree-level cross sections. The separate hard ($H$), jet ($J$), and soft ($S$) factors in Equation~(\ref{eq:factorization_formula}) do not contain large perturbative logarithms when evaluated at $\mu \sim \Lambda$, $\mu\sim \sqrt{\lambda}\Lambda$, and $\mu \sim \lambda \Lambda$, respectively. To control large logarithms, we renormalize to a common scale, and include terms enhanced by the emission of multiple photons~\cite{Yennie:1961ad,Mukhi:1982bk,Curci:1978kj,Smilga:1979uj}.

Our general exclusive observable, depicted in Fig.~\ref{fig:CCQE} and described by Equation~(\ref{eq:factorization_formula}), is defined to contain all photons that have energy below $\Delta E$ or are within angle $\Delta \theta$ of the charged lepton direction. We focus on two important cases relevant for neutrino experiments. First, for electron-flavor events, energetic collinear photons are reconstructed together with the electron. Thus the ``jet observable" applies, with appropriate choices of $\Delta E$ and $\Delta \theta$ (we will use $\Delta E = 10\,{\rm, MeV}$ and $\Delta\theta=10^\circ$ for illustration). Second, for muon flavor, collinear photons are only a small fraction of all photons above the soft-photon energy threshold (below permille level at $E_\nu=2\,{\rm GeV}$, cf. Fig.~4 of Ref.~\cite{Tomalak:2022xup}), both because the effective $\Delta\theta$ for muons in realistic detectors is smaller and because angles of typical photons are larger, $\sim {m_\ell}/{(E_\ell + m_\ell)}$. Thus for muon-flavor events, the formal limit $\Delta \theta \to 0$ is a good approximation, i.e., only soft photons with energy below $\Delta E$ are included in the observable (we will use $\Delta E=10\,{\rm MeV}$ for illustration).

\subsection*{Results for Flavor Ratios}

\begin{figure}[tbp]
 \centering
 \includegraphics[height=0.69\textwidth]{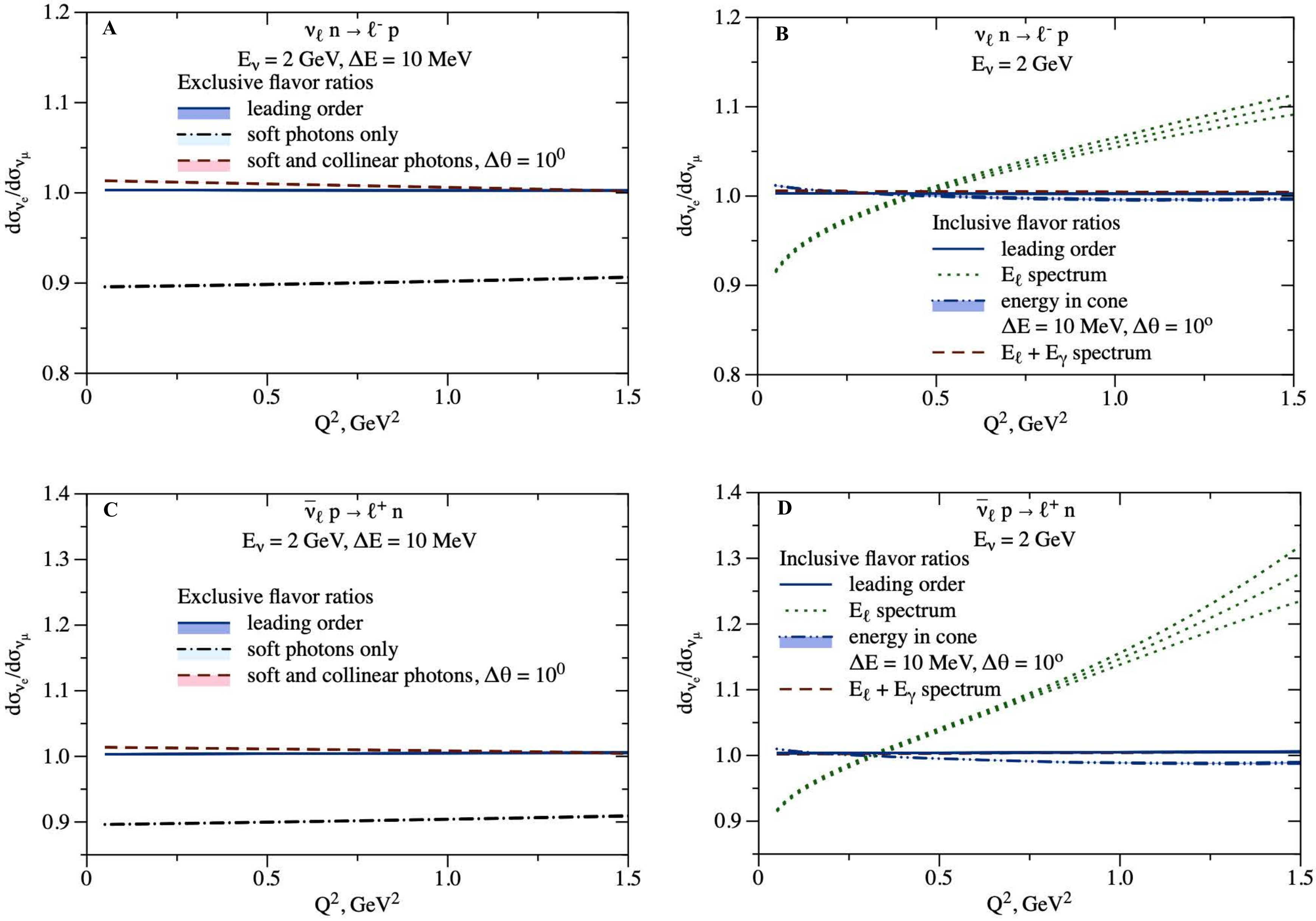}
 \caption{Radiatively corrected ratio of $\nu_e$ versus $\nu_\mu$ cross sections and corresponding uncertainty for exclusive observables ({\textbf{A}}, {\textbf{C}}) and inclusive observables ({\textbf{B}}, {\textbf{D}}). {\textbf{A}}, {\textbf{B}}: Neutrino scattering. {\textbf{C}}, {\textbf{D}}: Antineutrino scattering. For exclusive observables, the ratio of cross sections with soft photons of energy below $\Delta E = 10~\mathrm{MeV}$ is represented by the black dash-dotted lines. The tree-level ratio is shown by the blue solid lines. The red dashed lines with a legend ``soft and collinear photons" represent the ratio of jet observable (including photons in $\Delta \theta = 10^\circ$ cone for $\nu_e$ scattering) to the observable that excludes collinear photons for $\nu_\mu$ scattering. For inclusive observables, we display the $E_\ell$ spectrum (green dotted lines), $E_\ell+E_\gamma$ spectrum (red dashed lines), and ``energy in cone" spectrum (blue dash-double-dotted lines and filled band). For the ``energy in cone" spectrum, $Q^2$ is reconstructed from the energy in the $\Delta \theta = 10^\circ$ cone for electron flavor and from the lepton energy for muon flavor. \label{fig:xsection_ratio_muon_over_electron}}
 \end{figure}

Neutrino oscillation experiments aim to determine the relative flux of $\nu_e$ at a far detector originating from a primarily $\nu_\mu$ beam; this flux is interpreted as a $\nu_\mu \to \nu_e$ oscillation probability, and provides access to fundamental neutrino properties. The $\nu_e$ cross section is required to infer the flux of $\nu_e$ from observed event rates. Precise (anti)neutrino cross sections with electron flavor can be obtained from precise measurements of muon (anti)neutrino interactions at near detectors, combined with precise constraints on the ratio of electron and muon cross sections. Consequently, the electron-to-muon cross-section ratio is a critical ingredient in neutrino oscillation analyses~\cite{Day:2012gb,Coloma:2012ji,NuSTEC:2017hzk}.

We display this ratio in Fig.~\ref{fig:xsection_ratio_muon_over_electron}. For the exclusive case, we focus on our default observables with electron plus collinear and soft radiation, and muon plus soft-only radiation. For comparison, we also display the result when only soft radiation is included for the electron. In either case, dependence on hadronic physics is identical for $e$ and $\mu$ at the same value of hadronic momentum transfer, according to Equation~(\ref{eq:factorization_formula}), leaving only a small perturbative uncertainty on the ratio.

{\squeezetable
\begin{table}[hbp]
 \centering 
 \begin{tabular}{c|c|c|c|c|c|c}
 & \begin{tabular}{c}$E_\nu$, GeV\end{tabular} & & $\left(\frac{\sigma_e}{\sigma_\mu}-1\right)_\mathrm{LO}$, \% & $\frac{\sigma_e}{\sigma_\mu}-1$, \% \\ \hline
 \begin{tabular}{c}T2K/HyperK\end{tabular} & $0.6$ & \begin{tabular}{c} $\nu$ \\ $\bar{\nu}$\end{tabular} & \begin{tabular}{c} $ 2.47 \pm 0.06$ \\ $2.04 \pm 0.08$\end{tabular} & \begin{tabular}{c}$2.84 \pm 0.06\pm 0.37$\\$1.84 \pm 0.08 \pm 0.20$ \end{tabular}\\
 NOvA/DUNE & $2.0$ & \begin{tabular}{c} $\nu$ \\ $\bar{\nu}$\end{tabular} & \begin{tabular}{c} $0.322 \pm 0.006$ \\ $0.394 \pm 0.003$ \end{tabular} & \begin{tabular}{c} $ 0.54 \pm 0.01 \pm 0.22$ \\ $0.20 \pm 0.01 \pm 0.19$ \end{tabular}
 \end{tabular}
 \caption{ {\textbf{Flavor ratio of inclusive cross sections.}} Inclusive electron-to-muon cross-section ratios for neutrinos and antineutrinos without kinematic cuts. Uncertainties at leading order are from vector and axial nucleon form factors. For the final result, we include an additional hadronic uncertainty from the one-loop correction to the first uncertainty, and provide a second uncertainty as the magnitude of the radiative correction.}
 \label{tab:FlavorRatios}
\end{table}

As explained in more detail below, in addition to the exclusive case we consider inclusive observables that include all photon events in the cross section. For this case, we focus on the blue dash-double-dotted curve with the filled band in Fig.~\ref{fig:xsection_ratio_muon_over_electron}, corresponding to our default inclusive observables, i.e. including all photon events in the cross section, but reconstructing $Q^2$ using only collinear and soft radiation for the electron, and no radiation for the muon. For comparison, in Fig.~\ref{fig:xsection_ratio_muon_over_electron} we also display the results when both electron and muon events are reconstructed using only lepton energy ($E_\ell$ spectrum), and when both are reconstructed using all electromagnetic energy ($E_\ell+E_\gamma$ spectrum). Integrating over kinematics, we present the ratio of the total electron-to-muon cross sections for two kinematic setups without cuts on the lepton energy in Table~\ref{tab:FlavorRatios}.

\subsection*{Exclusive jet observables and impact of collinear photons}

\begin{figure}[tb]
 \centering
 \includegraphics[height=0.33\textwidth]{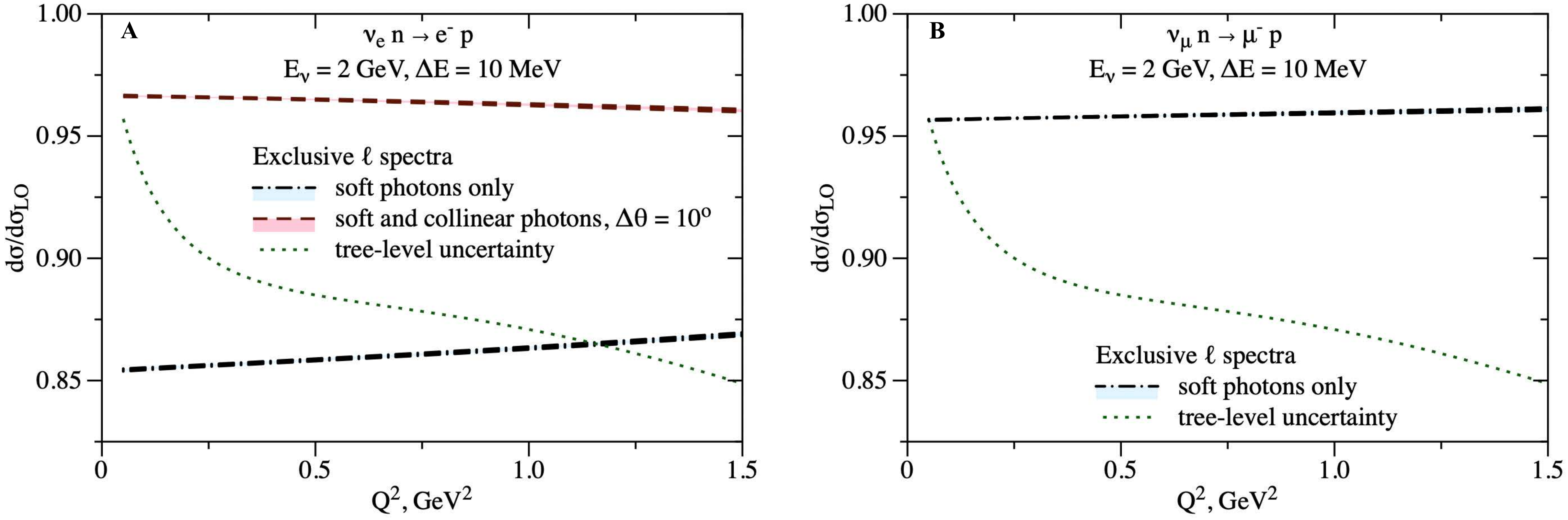}
 \caption{Radiatively corrected cross-section ratio to the tree-level result and corresponding uncertainty in neutrino scattering for exclusive observables. The legend corresponds to the panels {\textbf{A}} and {\textbf{C}} in Fig.~\ref{fig:xsection_ratio_muon_over_electron}. The tree-level uncertainty is represented by the green dotted line as a deviation of the ratio from unity. {\textbf{A}:} Electron flavor. {\textbf{B}:} Muon flavor. \label{fig:RC_n_excl}}
\end{figure}

The cross-section ratios for exclusive observables displayed in Fig.~\ref{fig:xsection_ratio_muon_over_electron} depend on whether collinear photons are included in the observable. Recall that while this specification depends in detail on detector capabilities and analysis strategies, our default observables are determined as follows: (1) soft radiation below $\Delta E$ is unobserved (but contributes to the cross section), independent of angle with respect to charged lepton direction; (2) collinear radiation accompanying electrons (within an angle $\Delta \theta$ of the electron direction) is included as part of the same electromagnetic shower; (3) collinear radiation accompanying muons is excluded. 

Fig.~\ref{fig:RC_n_excl} displays the ratio of the cross section to the leading-order (LO) result $\mathrm{d} \sigma_{\nu_\ell}/\mathrm{d} \sigma_\mathrm{LO}$, for default values $\Delta E = 10\,{\rm MeV}$ and $\Delta\theta =10^\circ$, as a function of nucleon momentum transfer $Q^2$. In the electron case, we compare our default jet observable (including energetic radiation within $10^\circ$ cone) to the soft-photon-only observable; the large correction $\sim 15\%$ in this case results from a logarithmic enhancement $\sim \ln(E_\nu/{m_e}) \ln(E_\nu/\Delta E)$. The factorization theorem of Equation~(\ref{eq:factorization_formula}) enforces a cancellation of hadronic uncertainty in the ratio of the corrected cross section to tree level, up to ${\cal O}(\alpha)$, resulting in the small uncertainty for the cross sections in Fig.~\ref{fig:RC_n_excl} (after the next-to-leading order resummation analysis, perturbative uncertainty is at or below permille level). For comparison, the plots also show the tree-level uncertainty on the cross section due to uncertain (dominantly axial-vector) nucleon form factors. This uncertainty cancels in the flavor ratios.

We remark that the ``soft photons only", dash-dotted curves in Fig.~\ref{fig:RC_n_excl}, are dramatically different for electrons and muons.  It is only after modifying the electron-neutrino cross section (by including also collinear photon radiation, the dashed curve on the left of Fig.~\ref{fig:RC_n_excl}) that it becomes similar to the muon-neutrino cross section (the dash-dotted curve on the right of Fig.~\ref{fig:RC_n_excl}).  There is an accidental coincidence of the $\sim 5\%$ corrections for the $\Delta\theta$-dependent electron-neutrino curve and for the $m_\mu$-dependent muon-neutrino curve.  This coincidence results in a ratio close to unity for the corresponding exclusive plots in Fig.~\ref{fig:xsection_ratio_muon_over_electron}.

\subsection*{Inclusive observables and impact of non-collinear photons}

The above ``exclusive" observables incorporate real photon radiation that is either unobservable by the detector (photon below $\Delta E$ in energy) or indistinguishable from the charged lepton (photon above $\Delta E$ in energy but within angle $\Delta \theta$ of the electron)~\cite{Bloch:1937pw,Yennie:1961ad,Kinoshita:1962ur,Lee:1964is}. Other hard photons are excluded from the cross section. However, oscillation experiments such as NOvA and DUNE that attempt to identify all neutrino charged-current interactions and determine neutrino energy by measuring the sum of lepton and recoil energy are likely to include such hard photon events.

To illustrate the impact of hard non-collinear photons  on kinematic reconstruction, we compute the spectrum with respect to several different choices for independent variable (``reconstructed $Q^2$"):
\begin{align}	
 Q^2_{\rm rec} &= 2 M \left( E_\nu - E_\ell - E_X \right), \,
\end{align}
where, for events without energetic photons, we have $E_X=0$; and, for events with an energetic photon of energy $E_\gamma$, we take (i) $E_X=0$ (``$E_\ell$ spectrum"); (ii) $E_X=E_\gamma$, when the photon is within $\Delta \theta=10^\circ$ of the electron, and $E_X=0$ otherwise (``energy in cone"); or (iii) $E_X=E_\gamma$ (``$E_\ell+E_\gamma$ spectrum").
\begin{figure}[tbp]
 \centering
  \includegraphics[height=0.33\textwidth]{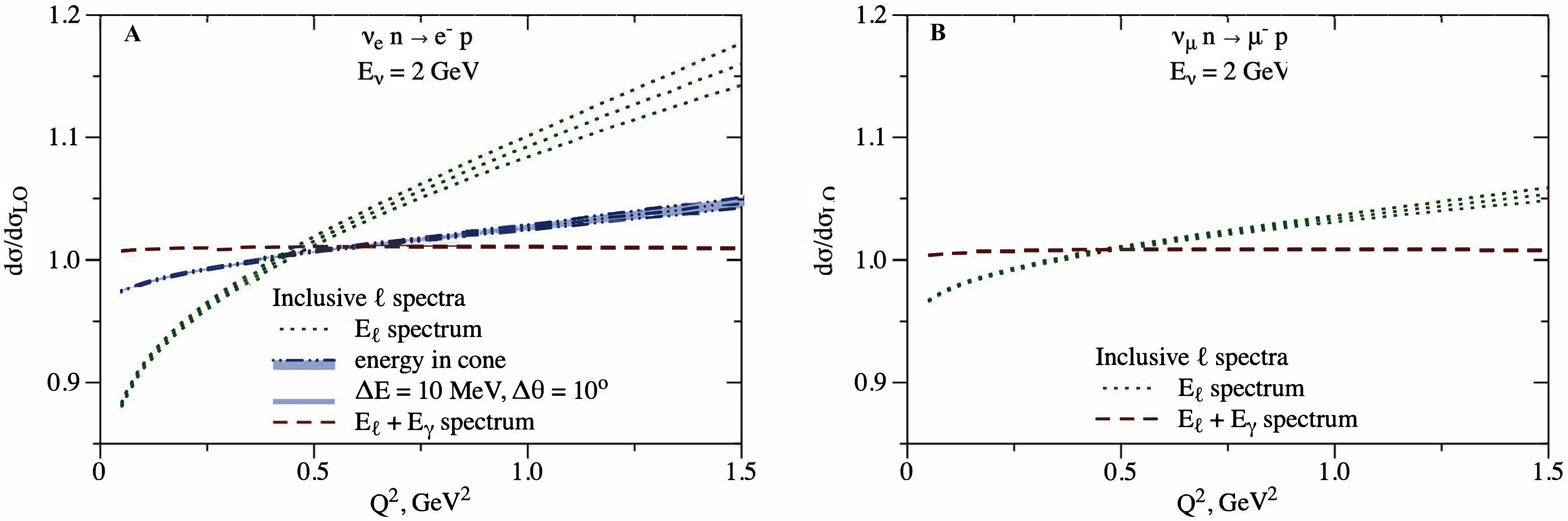}
\caption{Radiatively corrected cross-section ratio to the tree-level result and corresponding uncertainty in neutrino scattering for inclusive observables. The legend corresponds to the panels {\textbf{B}} and {\textbf{D}} in Fig.~\ref{fig:xsection_ratio_muon_over_electron}. {\textbf{A}:} Electron flavor. {\textbf{B}:} Muon flavor. \label{fig:RC_n_incl}}
\end{figure}
\begin{figure}[tbp]
 \centering
 \includegraphics[height=0.33\textwidth]{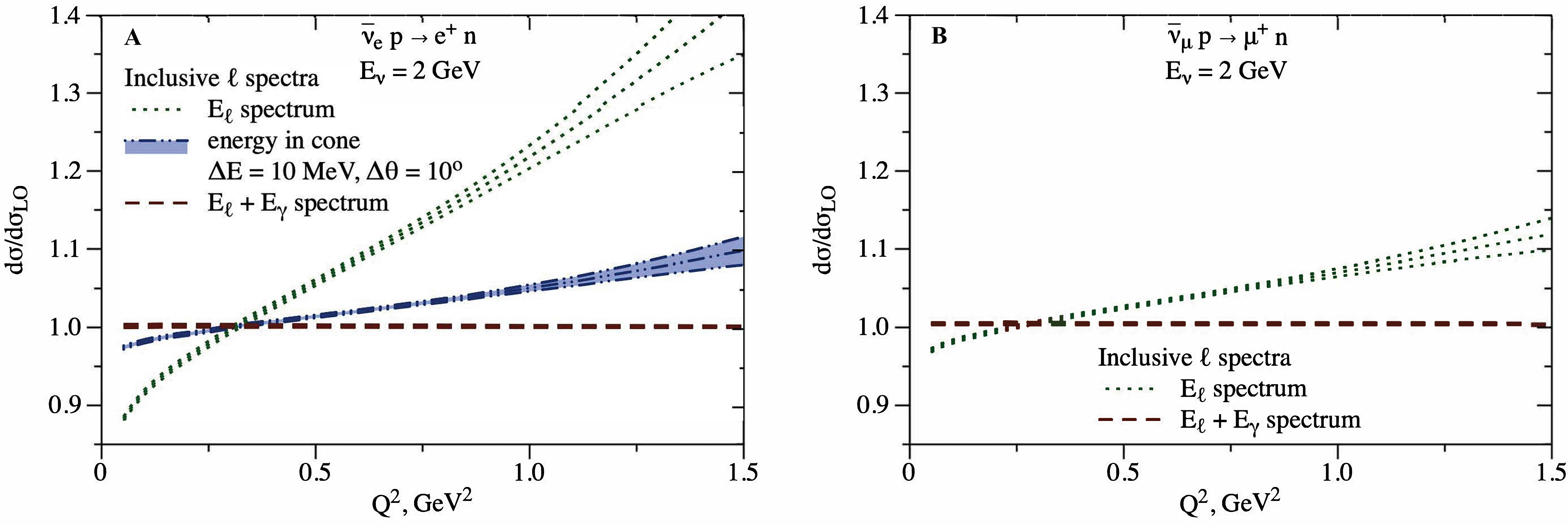}
\caption{Same as Fig.~\ref{fig:RC_n_incl} but for antineutrino scattering. \label{fig:RC_p_incl}}
\end{figure}

The results are displayed in Fig.~\ref{fig:RC_n_incl} for neutrino scattering and in Fig.~\ref{fig:RC_p_incl} for antineutrino scattering. There are several notable features of these curves. First, let us compare to the exclusive case displayed in Fig.~\ref{fig:RC_n_excl}. For electrons, the red dashed curves in the figures both represent spectra with respect to hadronic momentum transfer; the $\sim {\rm few}\,\%$ larger cross section in Fig.~\ref{fig:RC_n_incl} corresponds to the additional contribution from non-collinear energetic photons. Similarly for muons, the dash-dotted black curve in Fig.~\ref{fig:RC_n_excl} and the red dashed curve in Fig.~\ref{fig:RC_n_incl} both represent spectra with respect to hadronic momentum transfer, and their difference is identified with the contribution of energetic photons (of any angle). Second, although the three curves for $\nu_e$ in Figs.~\ref{fig:RC_n_incl} and~\ref{fig:RC_p_incl} (or two curves for $\nu_\mu$) integrate to the same total cross section, they differ significantly in their dependence on $Q^2_{\rm rec}$. It is essential to account for the correct kinematic dependence of radiative corrections when analysis cuts and acceptance effects are incorporated in practical experiments. For illustration, the curves in Figs.~\ref{fig:RC_n_incl} and~\ref{fig:RC_p_incl} integrate to cross sections that differ by up to~$10\%$ level (in this illustration the difference in partial cross sections computed with $Q^2<Q^2_{\rm cut}$ is divided by the total cross section, for different values of $Q^2_{\rm cut}$). Finally, as for the exclusive case, we remark that the directly comparable curves (green dotted ``$E_\ell$ spectrum") on the left and right of Figs. 4 and 5 are markedly different, and that the similarity of the blue ``energy in cone" curve on the left and the green ``$E_\ell$ spectrum" curve on the right results from an accidental cancellation involving the detector parameter $\Delta \theta$ and the lepton mass $m_\mu$.

\subsection*{Subleading and nuclear corrections}

We have used isospin symmetry to neglect isospin-violating tree-level form factors, to express results in terms of a common nucleon mass, and to obtain charged-current vector form factors from an isospin rotation of electromagnetic ones. Isospin-violating effects~\cite{Weinberg:1958ut,Behrends:1960nf,Dmitrasinovic:1995jt,Shiomi:1996np,Miller:1997ya,Govaerts:2000ps,Minamisono:2001cd,Kubis:2006cy,Lewis:2007qxa} due to electromagnetism are of order $ \alpha \approx 1/137$, and isospin-violating effects due to the quark mass difference $m_u-m_d$ are of order $\delta_N =(M_n - M_p)/M\approx 1.3\times 10^{-3}$ or $\delta_\pi =(m_{\pi^\pm}^2 -m_{\pi^0}^2)/m_\rho^2\approx 2.1\times 10^{-3}$, where $m_u, m_d, M_n, M_p, m_{\pi^\pm} ,m_{\pi^0}$ are masses of the up and down quarks, neutron and proton, charged and neutral pions, respectively; $m_\rho \approx 770\,{\rm MeV}$ is the $\rho$-meson mass representing a typical hadronic scale. In cross-section ratios to the tree-level results, $\mathrm{d}\sigma_{\nu_\ell}/ \mathrm{d} \sigma_\mathrm{LO}$, or in the ratio between lepton flavors, $\mathrm{d} \sigma_{\nu_e}/\mathrm{d} \sigma_{\nu_\mu}$, leading isospin-violating effects cancel, leaving corrections of order $\alpha \times \delta_{N, \pi} \sim 10^{-4}$ or $(m_\mu^2/M^2) \times \delta_{N, \pi} \sim 10^{-4}$. Hadronic uncertainties at leading and next-to-leading order in $\alpha$ are included in our analysis. Higher-order perturbative corrections are of order $\alpha^2 \sim 10^{-4}$. Power corrections are suppressed by $\Delta E/E_\nu$ or ${m_\mu^2/E_\nu^2}$, but enter at loop level and so are of order $\alpha \Delta E/E_\nu \sim \alpha {m_\mu^2/E_\nu^2} \sim 10^{-4}$.

Although the study was performed with (anti)neutrino-nucleon scattering, important cross-section ratios are insensitive to the explicit form of the nonperturbative hard function and similar conclusions are valid for scattering on nuclei. First, the radiative corrections to the exclusive cross sections in Fig.~\ref{fig:RC_n_excl} and the corresponding ratios in Fig.~\ref{fig:xsection_ratio_muon_over_electron} are dominated by large perturbative logarithms that are independent of nuclear or hadronic parameters. Second, for the inclusive cross sections displayed in Table~\ref{tab:FlavorRatios}, constraints on the lepton-mass dependence~\cite{Kinoshita:1962ur,Lee:1964is} imply small modifications to radiative corrections from nuclear effects. An explicit evaluation~\cite{Tomalak:2022xup} within the standard impulse approximation accounting for nucleon binding energy, initial-state Fermi motion, and final-state Pauli blocking yields corrections to $\sigma_e/\sigma_\mu$ of order $10^{-4}$ at $E_\nu=2\,{\rm GeV}$, and of order $10^{-3}$ at $E_\nu=0.6\,{\rm GeV}$, already contained in the hadronic error bars of Table~\ref{tab:FlavorRatios}.

\subsection*{Implications for Neutrino Oscillation Experiments}

The precise predictions for fully inclusive cross sections in Table~\ref{tab:FlavorRatios} have important implications for the T2K and NOvA experiments: the total cross section for electron-neutrino charged-current quasielastic (CCQE) events is precisely predicted in terms of observed muon-neutrino CCQE events. T2K and NOvA currently assume $2\%$ uncertainties on the extrapolation from muon (anti)neutrino to electron (anti)neutrino due to radiative corrections. In place of this assumption, our results provide a precise prediction, with reduced uncertainty. We also demonstrate that radiative correction uncertainty for both exclusive and inclusive observables can be controlled to the higher precision needed by the future DUNE and HyperK experiments. Before this work, the assumptions made by the current and future experiments were not justified by rigorous theoretical evaluation.

As Figs.~\ref{fig:xsection_ratio_muon_over_electron}, \ref{fig:RC_n_excl}, and \ref{fig:RC_n_incl} show, there can be large radiative corrections to the tree-level process: $\sim 15\%$ on the $\nu_e$ cross section, and $\sim 10\%$ on the muon-to-electron flavor ratio.  After introducing different definitions of the observable for electrons and muons, to conform to detector capabilities, the flavor ratio at the same kinematics (cf. Fig.~\ref{fig:xsection_ratio_muon_over_electron} left and Fig.~\ref{fig:RC_n_excl}) is remarkably close to unity; this is a consequence of an accidental cancellation involving the detector parameter $\Delta \theta$ for the electron, and the lepton mass $m_\mu$ for the muon. For total inclusive cross sections, a similar accidental cancellation happens (cf. Fig.~\ref{fig:xsection_ratio_muon_over_electron} right, Figs.~\ref{fig:RC_n_incl} and~\ref{fig:RC_p_incl}).

Differences between detection efficiency corrections and/or analysis cuts for electron and muon events can negate these cancellations in flavor ratios. In particular, experiments do not measure (anti)neutrino interactions in a way that is truly inclusive of final-state photons. The rate for events with non-collinear hard photons is between one percent and several percent of the total event rate, which is larger than the planned precision of future experiments. The experiments currently assume that non-collinear hard photons are absent, but such photons could disrupt event selection, particularly the separation of electrons from neutral pions or the exclusive identification of quasielastic events. Another effect of real photon radiation is the distortion of the reconstructed lepton energy spectrum, resulting in an enhancement of lower momentum leptons and depletion of higher momentum ones, cf. Figs.~\ref{fig:RC_n_incl} and~\ref{fig:RC_p_incl}. Because the inclusion of real photons is different for muon and electron reconstruction, this difference may change the relative efficiency of reconstructing the different neutrino flavors. Our results can be used to precisely account for these effects.

We note that our formalism can be used to address another important issue for modern neutrino oscillation experiments: when a muon from a charged-current $\nu_\mu$ interaction is accompanied by a sufficiently energetic collinear photon, the event can be misidentified as an electron charged-current interaction, confusing a particle identification algorithm looking for a penetrating muon track. Previous estimates for this effect~\cite{IwamotoThesis} were based on the splitting function approach of Refs.~\cite{DeRujula:1979grv,Day:2012gb}. The collinear approximation underlying the splitting function formalism is not a good approximation for the muon at GeV energies and it is important to revisit this question (the dimensionless parameter controlling collimation is not small; in fact, $\Delta \theta m_\mu /E_\nu$ is of order unity). We find that the probability of such muon misidentification is very small~\cite{Tomalak:2022xup}: less than a few$\times 10^{-4}$ for NOvA and DUNE, and less than $10^{-4}$ for T2K/HyperK.

\section*{DISCUSSION}

An important result from our studies for the precision accelerator neutrino oscillation program is that the total cross section as a function of (anti)neutrino energy, inclusive of real photon emission, is very similar for electron and muon (anti)neutrino events, as Figs.~\ref{fig:xsection_ratio_muon_over_electron}, \ref{fig:RC_n_incl}, and \ref{fig:RC_p_incl} illustrate. However, this simple result is achieved only after summing inclusively over distinct kinematical configurations. Electron-flavor and muon-flavor cross sections receive significant, and different, corrections as a function of kinematics that must be carefully accounted for when experimental cuts and efficiency corrections are applied in a practical experiment. It is also important to carefully match the theoretical calculation of radiative corrections to experimental conditions since radiative corrections depend strongly on the treatment of real photon radiation.

Current data on (anti)neutrino interactions do not have the precision to validate or challenge our precise calculations because of the sparse data on electron-neutrino and antineutrino scattering at these energies~\cite{Blietschau:1977mu,Abe:2014agb,Wolcott:2015hda,Abe:2015mxf,Abe:2020vot}. Experiments must therefore rely on this and other theoretical calculations to determine the effects of radiative corrections. Such effects can be potentially constrained by recent and forthcoming measurements with electrons~\cite{electronsforneutrinos:2020tbf,JeffersonLabHallATritium:2019xlj} and muons.

Applications to neutrino energy reconstruction, radiative corrections with pion and resonance production, and the inclusion of Coulomb and nuclear effects to general exclusive and inclusive observables, will be investigated in future work.

\section*{METHODS}

\subsection*{Hadronic model}

At tree level, the hard function appearing in Equation~(\ref{eq:factorization_formula}) can be conveniently expressed in terms of the structure-dependent quantities $A,B$, and $C$~\cite{LlewellynSmith:1971uhs}
\begin{equation}
H = \frac{\mathrm{G}^2_\mathrm{F} | V_{u d}|^2}{2\pi} \frac{M^2}{E_\nu^2} \left[ \left( \tau + r^2 \right)A(Q^2) - \nu B(Q^2) + \frac{\nu^2}{1+\tau} C(Q^2) \right] \,, \label{eq:xsection_neutrino}
\end{equation}
where $\tau = Q^2 / \left( 4 M^2 \right)$, $r = m_\ell/(2 M)$, $\nu = E_\nu/M - \tau - r^2$, $\mathrm{G}_\mathrm{F}$ is the Fermi coupling constant, and $V_{u d}$ is the Cabibbo-Kobayashi-Maskawa (CKM) matrix element. Assuming isospin symmetry, $A,~B$, and $C$ are expressed in terms of electric $G^V_E$, magnetic $G^V_M$, axial $F_A$, and pseudoscalar $F_P$, form factors as
\begin{align}
 A &= \tau \left( G^V_M \right)^2 - \left( G^V_E \right)^2 + (1+ \tau) F_A^2 - r^2 \left( \left( G^V_M \right)^2 + F_A^2 - 4 \tau F_P^2+ 4 F_A F_P \right) \,, \\
 B &= 4 \eta \tau F_A G^V_M \,, \\
 C &= \tau \left( G^V_M \right)^2 + \left( G^V_E \right)^2 + (1+ \tau) F_A^2 \,,\label{eq:ABC}
\end{align}
where $\eta=+1$ corresponds to neutrino scattering $\nu_\ell n \to \ell^- p$, and $\eta=-1$ corresponds to antineutrino scattering $\bar{\nu}_\ell p \to \ell^+ n$. In the evaluation of the hard function, we use form factors and uncertainties extracted from other data~\cite{Meyer:2016oeg,Borah:2020gte} for the tree-level contributions~\cite{LlewellynSmith:1971uhs}, and a gauge-invariant form-factor insertion model, which is motivated by~\cite{Maximon:2000hm,Blunden:2003sp,Graczyk:2013fha,Tomalak:2014dja,Tomalak:2022xup}, for the one-loop contributions. The form-factor insertion ansatz dresses point-particle Feynman diagrams with on-shell form factors at hadronic vertices. For the one-loop hard function, electromagnetic form factors are represented by dipoles with mass parameters varied as $\Lambda^2 \to (1\pm 0.1) \Lambda^2$ to cover the experimentally allowed range of form factors~\cite{A1:2013fsc,Borah:2020gte}. Uncertainties due to the insertion of on-shell hadronic vertices and the neglect of inelastic intermediate states are estimated by a simple ansatz that adds the neutron on-shell vertex to each of the neutron and proton electromagnetic vertices. Non-collinear hard photons introduce an additional hadronic structure beyond the hard function appearing in Equation~(\ref{eq:factorization_formula}). We estimate this effect by extending the form-factor insertion ansatz to describe real hard photon emission, employing the same gauge-invariant model as for the exclusive process. Uncertainties in the hard function largely cancel for the quantities presented in this paper, involving ratios of radiatively corrected and tree-level cross sections, or ratios of electron- and muon-flavor cross sections. Further discussion of the hadronic model for the hard function and its uncertainties are given in Ref.~\cite{Tomalak:2022xup}.

\subsection*{Soft and jet functions}

Here, we specify soft and jet functions from equation~(\ref{eq:factorization_formula}) at one-loop level. The process-independent soft function includes virtual corrections from the soft region and radiation of real soft photons below $\Delta E$. At one-loop level, the soft function is expressed as~\cite{Hill:2016gdf}
\begin{align}
 S\left( \frac{\mu}{\Delta E},\, v_\ell \cdot v_p,\, v\cdot v_\ell,\, v\cdot v_p \right) = 1 + \frac{\alpha}{\pi} \bigg[ 2 \bigg(1 - v_\ell\cdot v_p f(v_\ell\cdot v_p)\bigg) \ln \frac{\mu}{2 \Delta E} + G \left( v_\ell \cdot v_p, v\cdot v_\ell, v\cdot v_p \right) \bigg], \label{eq:soft_function}
\end{align}
where $v^\mu$ defines the laboratory frame in which $\Delta E$ is measured, $v_\ell^\mu$ and $v_p^\mu$ are the charged lepton and proton velocity vectors, and the functions $f$ and $G$ are given by~\cite{tHooft:1978jhc,Hill:2016gdf,Tomalak:2022xup}
\begin{align}
f(w) &= \frac{\ln w_+}{\sqrt{w^2-1}} \,, \\
G(w,\x,\y) &= \frac{\x }{ \sqrt{ \x^2-1} } \ln{\x_+} + \frac{\y }{\sqrt{ \y^2 -1}} \ln{y_+} + \frac{w}{ \sqrt{w^2-1}} \bigg[ \ln^2 \x_+ -\ln^2 \y_+ \nl
 &\quad + {\rm Li}_2\left( 1 - \frac{ \x_+ }{ \sqrt{w^2-1}} ( w_+ \x - \y ) \right) + {\rm Li}_2\left( 1 - \frac{ \x_- }{ \sqrt{w^2-1}} ( w_+ \x - \y ) \right) \nl
 &\quad - {\rm Li}_2\left( 1 - \frac{ \y_+ }{ \sqrt{w^2-1}} ( \x - w_- \y ) \right) - {\rm Li}_2\left( 1 - \frac{ \y_- }{ \sqrt{w^2-1}} ( \x - w_- \y ) \right) \bigg] \,,
\end{align}
with $a_\pm = a \pm \sqrt{a^2-1}$. 

The jet function includes virtual corrections from the collinear region and radiation of real photons within angle $\Delta \theta$ of the charged lepton direction. At one-loop level, the jet function is expressed as~\cite{Tomalak:2022xup}
\begin{align}
 J\left( \frac{\mu}{m_\ell}, \eta, x\right) = \left[ 1 + \frac{\alpha}{4 \pi} \left( \ln^2 \frac{\mu^2}{m_\ell^2} + \ln \frac{\mu^2}{m_\ell^2} + 4 + \frac{\pi^2}{6}\right) \right] \delta \left( 1 - x \right) + \frac{\alpha}{\pi} \left[ \frac{1}{2}\frac{1+x^2}{1-x}\ln(1+x^2\eta^2) -\frac{x}{ 1-x} \frac{x^2\eta^2 }{ 1 + {x^2\eta^2}} \right], \label{eq:jet_function_at_fixed_x}
\end{align}
with $\eta = \Delta \theta E_\ell / m_\ell$. The exclusive observables considered in this paper are given explicitly by integrating Equation~(\ref{eq:factorization_formula})
\begin{align}
\frac{\mathrm{d} \sigma}{\mathrm{d} Q^2} = \lim_{\epsilon\to 0} \int_{1-\epsilon}^{1+\epsilon} \frac{\mathrm{d} \sigma}{\mathrm{d} Q^2 \mathrm{d} x} \mathrm{d} x + \int_{0}^{1-\Delta E/E_\ell^{\rm tree}} \frac{\mathrm{d} \sigma}{\mathrm{d} Q^2 \mathrm{d} x} \mathrm{d} x \,,
\end{align}
where $E_\ell^{\rm tree}$ is the lepton energy for the tree-level process, and $x = E_\ell/E_\ell^{\rm tree}$ denotes the fraction of the total jet energy carried by the charged lepton (the total jet energy is defined as the energy carried by the charged lepton plus collinear photons).

Beginning at two-loop level, the factorization formula should be extended by the so-called remainder function~\cite{Hill:2016gdf,Tomalak:2022xup} that relates the running electromagnetic coupling in the QED theory with and without the dynamical charged lepton. We have suppressed this function for simplicity. Further details on higher-order perturbative corrections and resummation may be found in Ref.~\cite{Tomalak:2022xup}.

\subsection*{Photon energy cutoff and angular resolution parameters}

In this Section, we provide a simple estimate for the photon energy and angular acceptance parameters $\Delta E$ and $\Delta \theta$, using argon (with the nuclear electric charge $Z=18$) as the detector material and $E_e = 2\,{\rm GeV}$ as the electron energy for illustration. To determine $\Delta E$, we examine the different components of the total photon cross section in argon, and determine the energy at which $e^+e^-$ pair production starts to dominate over Compton scattering; this yields $\Delta E \approx 12~\mathrm{MeV}$~\cite{https://doi.org/10.18434/t48g6x}. To determine $\Delta\theta$, we consider the Moli\`{e}re radius of the electromagnetic shower initiated by the primary $e^\pm$ and the length of the mean shower maximum, and find the angle which would place the photon within the Moli\`{e}re radius at shower maximum. The Moli\`{e}re radius $R_M$ may be expressed as $R_M = X_0 E_s/E_c$~\cite{Nelson:1966jg,Bathow:1970dn,Zyla:2020zbs}, where $X_0$ is the radiation length, $E_s = \sqrt{(4\pi/\alpha)}m_e \approx 21~\mathrm{MeV}$, and $E_c$ is the critical energy of electrons, which we take in the form of the Rossi fit $E_c = 610/(Z+1.24)~\mathrm{MeV}$. The electromagnetic shower maximum length $L_M$ depends logarithmically on the electron energy $E_e$~\cite{Zyla:2020zbs} : $L_M \approx X_0 \left[ \ln( E_e/E_c) - 1/2 \right]$. The angular resolution parameter is thus:
\begin{align}
 \Delta \theta \approx \arctan \left( \frac{R_M}{L_M} \right). \label{eq:angular_resolution}
\end{align}
For $Z=18$ and $E_e=2\,{\rm GeV}$, we find $\Delta \theta \approx 10^\circ$.

\section*{Data availability}

The cross-section data generated and used in this study are provided in the Supplementary Information file.

\section*{Code availability}

The code to reproduce all plots and results in this study is provided in the Supplementary Information file.

\bibliography{radcorr}

\section*{Acknowledgments}

We thank Clarence Wret for checking the leading order calculations presented here against generator calculations, Kaushik Borah for an independent validation of antineutrino-proton cross-section expression, Emanuele Mereghetti and Ryan Plestid for discussions. This work was supported by the U.S. Department of Energy, Office of Science, Office of High Energy Physics, under Awards DE-SC0019095 and DE-SC0008475. Fermilab is operated by Fermi Research Alliance, LLC under Contract No. DE-AC02-07CH11359 with the United States Department of Energy. O.T. acknowledges support by the Visiting Scholars Award Program of the Universities Research Association, theory groups at Fermilab and Institute for Nuclear Physics at JGU Mainz for warm hospitality. O.T. is supported by the US Department of Energy through the Los Alamos National Laboratory. Los Alamos National Laboratory is operated by Triad National Security, LLC, for the National Nuclear Security Administration of U.S. Department of Energy (Contract No. 89233218CNA000001). This research is funded by LANL's Laboratory Directed Research and Development (LDRD/PRD) program under project number 20210968PRD4. Q.C. acknowledges KITP Graduate Fellow program supported by the Heising-Simons Foundation, the Simons Foundation, and National Science Foundation Grant No. NSF PHY-1748958. R.J.H. acknowledges support from the Neutrino Theory Network at Fermilab. K.S.M. acknowledges support from a Fermilab Intensity Frontier Fellowship during the early stages of this work, and from the University of Rochester's Steven Chu Professorship in Physics.

\section*{Author contributions}

O.T., Q.C., R.H., and K.S.M. contributed substantially to the results and to the writing of the manuscript. The initial factorization calculation and the numerical evaluations for the plots and tables in the paper were performed by O.T.

\section*{Competing interests}

The authors declare no competing interests.

\end{document}